\documentclass[twocolumn,showpacs,preprintnumbers,amsmath,amssymb]{revtex4}
\usepackage{graphicx}
\usepackage{dcolumn}
\usepackage{bm}

\begin{document}

\title{Periodic forcing in a three level cellular automata model for a vector transmitted disease}
\author {L. B. L. Santos}
\affiliation{Instituto de F\'{\i}sica, Universidade Federal da
Bahia, 40210-340, Salvador, Brazil}
\author  {M. C. Costa}
\affiliation{Instituto de F\'{\i}sica, Universidade Federal da
Bahia, 40210-340, Salvador, Brazil}
\author  {S. T. R. Pinho}
\altaffiliation[] {Corresponding author, email: suani@ufba.br }
\affiliation{Instituto de F\'{\i}sica, Universidade Federal da Bahia,
40210-340, Salvador, Brazil} \email[correspondent author]{suani@ufba.br}
\thanks{corresponding author}
\author  {R. F. S. Andrade}
\affiliation{Instituto de F\'{\i}sica, Universidade Federal da
Bahia, 40210-340, Salvador, Brazil}
\author  {F. R. Barreto}
\affiliation{Instituto de Sa\'ude Coletiva, Universidade Federal da
Bahia, 40110-140, Salvador, Brazil}
\author  {M. G. Teixeira}
\affiliation{Instituto de Sa\'ude Coletiva, Universidade Federal da
Bahia, 40110-140, Salvador, Brazil}
\author {M. L. Barreto}
\affiliation{Instituto de Sa\'ude Coletiva, Universidade Federal da
Bahia, 40110-140, Salvador, Brazil}
\date{ \today }

\begin{abstract}

The transmission of vector infectious diseases, which produces complex
spatiotemporal patterns, is analyzed by a periodically forced
two-dimensional cellular automata model. The system, which comprises three
population levels, is introduced to describe complex features of the
dynamics of the vector transmitted dengue epidemics, known to be very
sensitive to seasonal variables. The three coupled levels represent the
human, the adult and immature vector populations. The dynamics includes
external seasonality forcing (rainfall intensity data), human and mosquito
mobility, and vector control effects. The model parameters, even if
bounded to well defined intervals obtained from reported data, can be
selected to reproduce specific epidemic outbursts. In the current study,
explicit results are obtained by comparison with actual data retrieved
from the time-series of dengue epidemics in two cities in Brazil. The
results show fluctuations that are not captured by mean-field models. It
also reveals the qualitative behavior of the spatiotemporal patterns of
the epidemics. In the extreme situation of absence of external periodic
drive, the model predicts completely distinct long time evolution. The
model is robust in the sense that it is able to reproduce the time series
of dengue epidemics of different cities, provided the forcing term takes
into account the local rainfall modulation. Finally, the dependence
between epidemics threshold and vector control undergoes a transition from
power law to stretched exponential behavior due to human mobility effect.
\end{abstract}



\smallskip\

\maketitle

\section{INTRODUCTION}
Understanding the rather complex dynamics of transmissible diseases
is of utmost importance for improving life quality, and even the
survival of some human population groups. To achieve this,
interdisciplinary efforts are necessary, which certainly include the
use of the recently techniques developed to study complex systems
\cite{complex1,complex2,May}. At the beginning of 21 century, both
directly transmitted diseases, like tuberculosis and AIDS, as well
as vector-transmitted diseases, such as dengue and malaria, are
still not controlled. In modern life, the intense flux of people at
global level and within large cities \cite{Chowell} increases the
complexity of the propagation of transmitted diseases
\cite{Gubler1}. For vector-transmitted diseases, there are already
indications that climatic conditions and vector mobility may
increase the number of cases \cite{Hales}. In the case of dengue, an
arboviral disease transmitted to humans by {\it Aedes} mosquitoes
(mainly {\it Aedes Aegypti}), several determinant factors for its
transmission are found in large urban centers \cite{Kuno}: human
concentration, large inter- and intra-city human mobility, the
climatic conditions for the vector proliferation (high humidity and
temperature between $15^oC$ and $40^oC$). Accordingly, it is found
that the dengue outbursts are quite sensitive to seasonal variations
in pluviometric precipitations, humidity and temperature. The
disease, which may be caused by four different virus serotype
(DenV1-DenV4), reaches yearly some 50 millions people in more than
60 countries, with $\sim$ 21000 casualties \cite{WHO}.

Since 1992 \cite{Newton}, ordinary differential equation (ODE)
models have been proposed to analyze dengue inter-host dynamics and
the effect of vector control actions. More recently, some attempts
to introduce the spatial dependence on the disease propagation have
been reported, using both partial differential equation (PDE)
\cite{Maidana} and cellular automata (CA)\cite{Ferreira} models, and
other data analysis techniques \cite{Vecchio}. In \cite{Ferreira},
the authors proposes a model that takes into account only the
description of mosquito population, which may be found in the adult
phase, and the immature phase comprising several stages. However, a
more accurate description of the dengue propagation must include,
besides the interaction among these population groups, the vector
mobility, effect of control actions, and an explicit climatic
periodic forcing on the population variables. To our knowledge, no
previous investigation has taken into account all of these factors.

In this work, we investigate an inter-host three level CA model,
which describe  the pertinent population groups in a urban
environment: human, adult vector mosquito, and immature vector in
the aquatic phase. As we will detail later on, it includes all of
the quoted effects: external forcing to describe the environment
influence on the vector life cycle, as well as other interaction
terms describe the effect of human and vector mobility and control
actions. The results provided by the model reproduce actual time
series from some well document dengue epidemics in specific years
urban centers in Brazil. Besides that, they also qualitatively agree
with main features of the spatiotemporal patterns. We also show
that, in the absence of a periodic forcing, the actual epidemic
outbursts are not reproduced, supporting the claims of the
importance of climatic aspects in the triggering of local events.
Finally, as the model describes the behavior of the exposed
population for larger time intervals under the presence of climatic
seasonal variations, it is possible to follow the effect of vector
control actions. In such case, our results indicate a power law
dependence between the epidemic threshold and the parameter
describing the intensity of vector control.

The current description of vector-transmitted diseases goes along
several successful works based on CA intra-host disease propagation
models (for instance, AIDS, \cite{HIV}, malaria \cite{malaria},
cancer \cite{cancer}) and also on inter-host models
\cite{interhost}. It is also worth mentioning that the presence of
multiple CA interacting levels in epidemic models has been explored
in alternative topologies, as that of complex networks where nodes
represent patches of regular lattices \cite{Lobato1} submitted to a
contact process dynamics \cite{Lobato2}.

The paper is organized as follows: in Section \ref{sec2}, we
introduce the CA local rules, comparing them to other models in the
literature. Section \ref{sec3} discusses the choice of parameter
values in our simulations. In Section \ref{sec4}, we present our
results, comparing them with actual data: the simulated time series
(\ref{subsec41}) resulting from the periodic forcing seasonal
effects, the simulated spatiotemporal patterns and the vector
control associated to human mobility effect. Finally, Section
\ref{sec5} closes the paper with concluding remarks and
perspectives.

\section{The model}
\label{sec2} Some of the basic interaction mechanisms and external effects
be included in our three level CA model have been used, in other context,
by previous ODE models reported in the literature. The first attempt
\cite{Newton} considered a compartment model, in which humans follow SEIR
(susceptible, exposed, infected and removed) dynamics. Since mosquitoes
usually die before being removed, the authors consider that they follow a
simpler three-compartment SEI version.  On the other hand, climatic effects
were modeled by seasonal variations of model parameters by an ODE system
\cite{Coutinho}. Tuning models by comparison to actual data have also been
attempted, e.g., by the estimation of the basal transmission rate for
age-stratified data from Thailand \cite{Fergunson}. Other models have
considered the role of a unique vector in the transmission of multiple
diseases, as more than one dengue serotype \cite{Esteva,Bartley,Schwartz}
or the concurrent transmission of yellow fever in dengue infested areas
\cite{Massad}. Finally, the effect of vector control have already been
explicitly analyzed in ODE models \cite{Bartley,esteva_yang,Schwartz}.

\begin{figure}[!h]
\begin{center}
\includegraphics*[width=5cm]{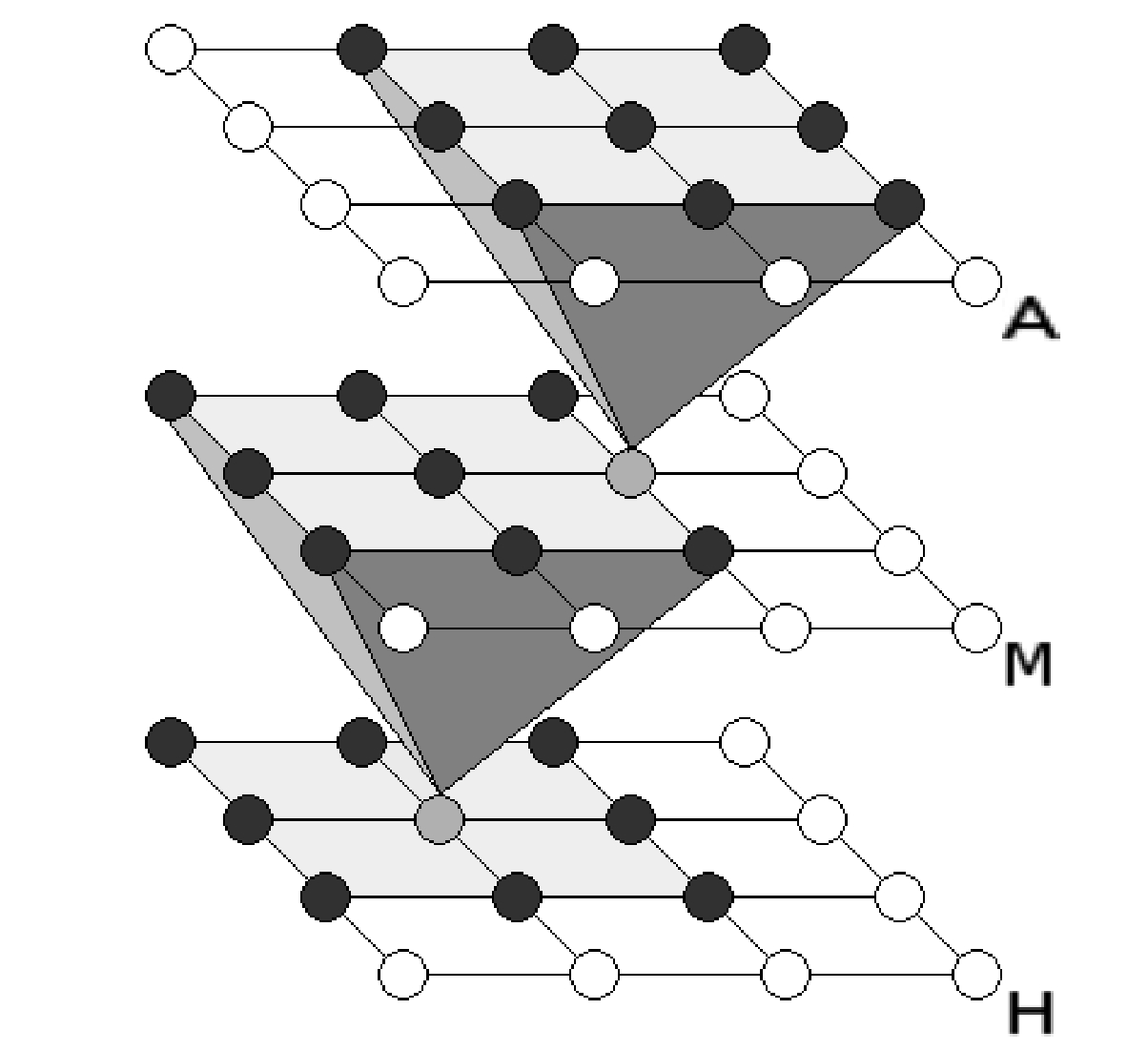}
\end{center}
\caption{\label{Figure1} Diagram of different lattices: humans (\emph{H}),
mosquitoes (\emph{M}) and aquatic phase (\emph{A}). Note that each element
of lattices \emph{H} and \emph{A} `sees' up to nine neighbors of the
lattice \emph{M} (and vice-versa).}
\end{figure}

Each of three CA levels consists of a two-dimensional square lattice with
$N_s=L \times L$ sites. Correspondingly, the CA is subjected to closed
boundary conditions {because it mimics dengue transmission in a city}. If
we compare the results to actual data, each neighborhood corresponds to a
set of distinct spatial units (census sectors) into which the reported
cases are assigned to. Each site in the distinct levels describe,
respectively, the local populations: human ($H$), mosquito ($M$) and
immature vector in the aquatic phase ($A$). The CA inter-layer interaction
rules couple, locally, the three involved levels due to the interactions
between $H$ and $M$ levels, and the $A$ to $M$ flux of the vector
population. The CA Moore neighborhood with radius 1 allows, for each site
of a given layer, a maximum of 9 neighbors in the level it interacts with
(see Figure \ref{Figure1}). We restrict ourselves to the one-serotype
situation, although the model can be extended to simulate the dynamics
with more than one serotype.

\begin{figure}[!h]
\begin{center}
\includegraphics[width=9.0cm,angle=0]{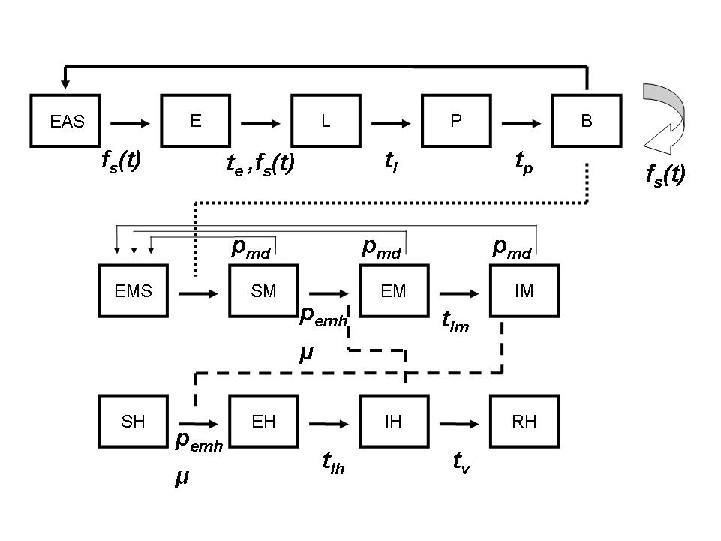}
\end{center}
\caption{\label{Figure2} A schematic representation of the local
rules of the model.}
\end{figure}

According to previously indicated models, in the $A$ phase, the
vector is found in one of 4 compartments: egg ($E$), larvae ($L$),
pupae ($P$) and breeding ($B$). The $M$ phase comprises 3
compartments: susceptible ($SM$), exposed ($EM$), and infectious
($IM$). Finally, considering only one serotype, there are 4 possible
compartments for $H$ sites: susceptible ($SH$), exposed ($EH$),
infectious ($IH$), and recovered ($RH$). Moreover, sites of $A$ and
$M$ levels can be in empty states, denoted by $EAS$ and $EMS$. The
local interaction rules, based on the entomological \cite{vector}
and epidemiological aspects \cite{Teixeira}, are such that, for each
level: (see Figure \ref{Figure2}).

\begin{itemize}
\item []

{\bf $A$ level:} $E$, $L$, $P$ and $B$ states evolve from the
preceding one after the $E$ eclosion period $t_e$, $L$ phase period
$t_l$ and $P$ phase period $t_p$. An empty site $EIM$ may be
replaced with probability $f_s(t)$ by an $E$ state, if there is at
least an occupied site in its Moore neighborhood at the $M$ level.
The transition from $E$ to $L$ compartments also depends on
$f_s(t)$, much as the persistence of $B$, which releases an adult
mosquito $SM$ to a $EPM$ site of the $M$ level.

\vspace{0.4cm}

\item []

{\bf $M$ level:} The population in the $M$ level results from the dynamics
in $A$ phase. Adult population $M$ dies according to a death probability
$p_{dm}$ in any state. The transition from a $SM$ site into $EM$ depends
on the number of $IH$ sites in its neighborhood in the $H$ level, on the
local effective biting humans-mosquitoes probability $p_{ehm}$, and on the
human mobility $\mu$. An $EM$ site becomes $IM$ after the $M$ virus latent
period $t_{lm}$.

\vspace{0.4cm}

\item []

{\bf $H$ level:} In a similar way to the $SM\rightarrow EM$ transition, a
$SH$ site changes to $EH$ according to the local effective
mosquitoes-humans biting probability ($p_{emh}$), the number of $IM$ sites
in the $M$ level neighborhood, and on the human mobility $\mu$. $EH$
becomes infectious $IH$ after the $H$ virus latent period $t_{lh}$, and
$IH$ becomes recovered $RH$ after the viremia period $t_v$.

\end{itemize}

Note that, in the above level descriptions, we already included
relevant features of dengue transmission that we have called the
attention in Section 1. Seasonal information (rainfall intensity) is
used as input data \cite{Hales} \cite{Kuno} by tuning the time
dependence of the $f_s(t)$ probability, using a Fourier expansion of
the actual rainfall series.  If the time series do not include daily
entries, or is not complete over the whole simulation period,
interpolation or addition of random noise to the day average taken
over a few years can be used. Global infection probabilities between
$H$ and $M$ populations, due to mobility in private and public
transport systems, is described by a global (mean-field) mobility
parameter $\mu$. The action of $\mu$, which is the same for all
sites, is to globally increase the $SM\rightarrow EM$ and
$SH\rightarrow EH$ probability transitions, without any influence
from the neighborhood population in the other level.

Finally, the decrease of populations in  $M$ level resulting from vector
control actions is included by the following additional rule: the natural
$M$ death probability is increased by an additional amount $p_{adm}$,
which reduces the adult mosquitoes on any state of the M level.

\section{Parameter values}
\label{sec3} The CA parameters introduced in the previous section
can be classified into four classes, according to the individual
process they describe: 1) Spatial parameters, as $L$ and $\mu$; 2)
Temporal parameters: $t_e$, $t_l$, $t_p$, $t_{lm}$, $t_v$, and
$t_{lh}$; 3) The probabilities of transmission and mosquito death
parameters: $p_{emh}$, $p_{ehm}$, and $p_{md}$; 4) Vector control
parameter: $p_{adm}$.

\begin{table*}[!h]
\begin{center}
\begin{ruledtabular}
\begin{tabular}{|c|c|}
\hline $Parameter$  & {\it Range of values} \\
\hline Egg period ($t_e$) \cite{vector}  & 4-5 days \\
\hline Larvae phase period ($t_l$) \cite{vector}  & 5-7 days \\
\hline Pupae phase period ($t_p$) \cite{vector}  & 2-3 days \\ \hline
Latent period of virus in the mosquito ($t_{lm}$) \cite{Yang,Newton,vector,Siler} & 7-20 days \\
\hline Latent period of virus in the human ($t_{lh}$) \cite{Newton,Yang,Siler}  & 2-12 days \\
\hline Viremia period ($t_v$) \cite{Newton,Yang,Gubler2}  & 3-7 days \\
\hline Probability of transmission human-mosquito ($p_{ehm}$) \cite{Rosen}  & 0.5-1.0 \\
\hline Probability of transmission
mosquito-human ($p_{emh}$) \cite{Newton}  & 0.5-1.0 \\
\hline Probability of mosquito death ($p_{md}$) \cite{Newton,Massad,Sheppard} & 0.128-0.25 \\
\hline
\end{tabular}
\end{ruledtabular}
\end{center}
\caption{\label{Table1}The parameter range of values of temporal
parameters and the probabilities of transmission H-M and M-H, and of
death mosquito according to the literature. The baseline values were
chosen for the simulations of the model.}
\end{table*}

The values of spatial parameters are obtained by taking into account
the data of a given urban center. We estimate the size $L$ of the
lattice (number of sites = $L^2$) by the area of the city ($A_c$)
and the flight radius of the vector ($R$). More specifically, we
assume that $A_c=L^2 a$, where $a$ is the area of one cell, while
$R$ corresponds to the average {\it(Moore)} neighborhood radius.
This way, we have

\begin{equation}
\label{Lestimation} R=\frac{\sqrt{a} (1+\sqrt{2})}{2}
\Longrightarrow L=\sqrt{\frac{A_c}{a}}=\frac{\sqrt{A_c}
(1+\sqrt{2})}{2R}
\end{equation}

As the dispersion of {\it Aedes aegypti} due to its flight rarely
exceeds $100m$ \cite{Liew}, we assume $R=100m$. The range of values
of $\mu$ was estimated by requiring that the model reproduces the
same behavior of the histogram of the number of census sectors with,
at least, one reported dengue case during the corresponding time
period.

We assumed fixed values (within the range presented in Table \ref{Table1})
for the probabilities of transmission $p_{ehm}=p_{emh}=0.75$ \cite{Newton}
and of mosquito death $p_{md}=1/7=0.143$. For vector control parameter,
when is the case, we scrutinize the complete interval from 0 to 1.

Choosing the CA iteration time unit to be one day, we are able to set
value intervals for several temporal parameters according to the
literature (see Table \ref{Table1}). To obtain baseline values for
temporal parameters and epidemic threshold, we adapt the epidemiological
definition of an epidemic process \cite{epidemiobook} to our model
simulations. A disease is considered epidemics if the annual incidence
$I$,  the number of reported case to susceptible population, is above a
certain (epidemics) threshold $I_{th}$. Therefore, $I_{th}$ may be given
by
\begin{equation}
\label{epidemics} I_{th} = <I> + 2 \sigma,
\end{equation}
where the average incidence $<I>$ is calculated with respect to the last
$N$ years and $\sigma$ corresponds to the standard deviation. To obtain
corresponding model values, we run the program for $N$ different random
seeds. We recall that, as for actual cases of vector transmitted diseases,
several numerical simulations resulting from different random seeds die
out in the first weeks, being characterized as small endemic processes.

After the evaluation of $<I>$ and $\sigma$, we run the program as
many times as necessary to get $K$ independent samples with
$I>I_{th}$. Although we perform the numerical simulations of the
model for large time intervals, our analysis can be restricted to
364 time unit intervals if we want to compare the results with
actual data of one year epidemics series. The output data are the
time series of density of each state in the $H, M$ and $A$ levels of
the CA model, and the spatiotemporal configurations at any time
step. The cpu time increases according to $L^3$ and linearly with
the number of samples.

Finally, based on the range of values in Table \ref{Table1} for
temporal parameters, simulations have been conducted for an initial
set of parameter values. Then, we investigate the effect of changing
one by one parameter, while holding all the others fixed. This way,
we identify the baseline values that minimizes the error between the
actual time series and the simulated time series. We perform several
tests in order to check the robustness of the chosen initial set of
parameter values. For a systematic analysis of parameter values, we
considered an average of $M$ simulations samples, identifying the
best output for the purpose of comparison with one actual epidemics
time series. This is achieved by the analysis of the minimum
discrepancy between actual and simulated time series:

\begin{equation}
\label{error} e= \frac{\sum_{i=1}^{T} |a_i-s_i|}{T},
\end{equation}
where $T$ is the number of days, $a_i$ is the actual incidence and
$s_i$ is the simulated incidence of day $i$.

Once estimated the baseline of temporal parameter values, the
analysis of minimal discrepancy is also applied to select the best
sample in comparison to actual data.

\section{RESULTS}
\label{sec4} In order to validate the model, we consider the data of
the first dengue epidemics (DenV-2) in 1995, Salvador, Brazil
\cite{Barreto}, when its population $p_c=2.3$ million habitants
distributed over an $A_c=313 \times 10^6 m^2$ area. In 1995, the
average daily temperature was $25.89 ^oC$ with $1.47$ standard
deviation. The city yearly mean precipitation is 1980 mm/year, while
seasonal effects concentrate precipitation in the months
March-August.

\subsection{The seasonal effects: actual and simulated time series}
\label{subsec41}

\begin{figure}[!h]
\begin{center}
\includegraphics*[width=7.0cm]{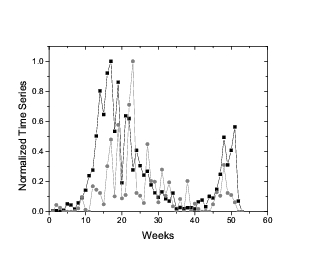}
\end{center}
\caption{\label{Figure3} Normalized time series of weekly $I_D$ in
Salvador (black-squares) and normalized time series of weekly $\Gamma_R$
(grey-circles) for 1995. The normalization factor are 846 cases and 373 mm
respectively.}
\end{figure}

The 1995 weekly rain intensity $\Gamma_R$ and reported number of new
dengue cases $I_D$ (incidence) are shown in Figure \ref{Figure3}, where
the data have been normalized by the largest input for the sake of
comparing the tendency of the curves. As, in this case, temperature and
humidity are quite stable, rainfall is the most important climatical
factor for dengue propagation. Indeed, the Pearson correlation varies from
0.49 to 0.76 for, respectively, weekly and monthly sampled data. As it
will be clear from the discussion of our results, such increase in the
correlation in value is due to a roughly two week delay time between the
two signals. When the series are clustered in large time windows, such
effects become much smaller. The daily rainfall data was provided by the
Brazilian government \cite{INMET}.

The simulations are based on the function $f_s(t)$ corresponds to
the Fourier expansion
\begin{eqnarray}
f_s(t) & = & a_0 + \sum_{j=1}^{12} a_j\, cos( \pi j t/26) + b_j \,
sin( \pi j t/26)
\end{eqnarray}
with $a_0=0.13585$, $a_1=-0.12872$, $b_1=0.05071$, $a_2=0.0502$,
$b_2=-0.0882$, $a_{12}= 0.00744$, $b_{12}=0.04713$. The total contribution of the
the remaining coefficients $a_i,b_i$ can be
neglected.

The $10,831$ reported dengue cases in Salvador during 1995 were
geo-referenced by epidemiological week (52 temporal units) and census
sectors (2600 spatial units)\cite{Barreto}. Note that, due to large
official sub-notifications (26 \%), the actual number of cases is much
larger. In Salvador, the epidemics peak occurs before the rainfall peak,
what can be justified by the fact that, due to the high intense
pluviometric precipitation peak, the rainfall washes out the vector in the
immature phase. As well will see later, this may not happen in other urban
centers. According to expression (\ref{Lestimation}), we are lead to the
value $L=214$. On the other hand, the value $\mu =5 \times 10^{-4}$, has
been selected from the interval where the model is able to reproduce the
exponential behavior in the probability distribution of observed new cases
in a year among 2600 sensus sectors (not shown).

\begin{figure}[!h]
\begin{center}
\includegraphics*[width=7.0cm]{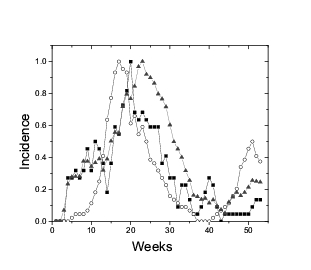}
\end{center}
\caption{\label{Figure4} Actual ($I_D$) and simulated ($IH_N$) weekly
incidence time series of Salvador in 1995 normalized by largest single
input. The data has been smoothed by averaging on three consecutive weeks.
Circles, squares and triangles indicate, respectively, $I_D$, the best
individual sample, and average value over 20 samples taken from random
seeds. The normalization factors for $I_D$, the best $IH_N$ and the
averaged $IH_N$, are 17, 11 and 14 respectively. Consider one sample and
the following parameter values: $t_e=5$, $t_l=5$, $t_p=3$ , $t_{lm}=7$,
$t_{lh}=6$, $t_v=6$, $p_{ehm}=p_{emh}=0.75$, $p_{md}=0.143$, $\mu=5 \times
10^{-4}$.}
\end{figure}

Assuming that there is one infected individual in each site of $H$
lattice, the best sample is able to reproduce the actual data quite well,
as shown by the normalized actual and simulated incidence time series in
Figure \ref{Figure4}. We normalize both the actual and simulated time
series for the purpose of avoiding distortions due to large
sub-notifications. To set up the correspondence between $I_R$ and the
simulated incidence $IH_N$, that is, the number of new infected humans at
a time step, we use the scale factor $L^2/p_c$. The normalization factors
for the actual and the best simulated incidence time series result,
respectively 17 and 11.

Note that the delay between the peaks of $I_D$ and the best individual
sample is much smaller that the delay between $\Gamma_R$ and $I_D$ in
Figure \ref{Figure3}, even considering the averaging on three consecutive
weeks which amplify the delay effect. Although this effect is also
amplified for the average over some samples, it is still smaller than the
delay between $I_D$ and $\Gamma_R$ in Figure \ref{Figure3}.

\begin{figure}[!h]
\begin{center}
\includegraphics*[width=7.0cm]{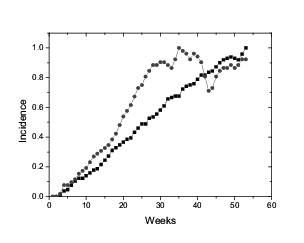}
\end{center}
\caption{\label{Figure5} Normalized average over 20 samples of simulated
weekly incidence ($IH_N$) when $f_s(t)=1$ (black square) and
$f_s(t)=sin(2\pi t/52)$ (grey circle). The normalization factors for
$f_s(t)=1$ and $f_s(t)=sin(2\pi t/52)$ are 35.5 and 17.3 respectively. The
data has been smoothed by averaging on three consecutive weeks. Parameter
values are the same as in Figure \ref{Figure4}.}
\end{figure}

To emphasize the importance of the periodic forcing to recover the
reported $I_D$ values, we draw, in Figure \ref{Figure5}, the time
evolution according to two hypothetical scenarios. They were
obtained by replacing $f_s(t)$, in first place, by a constant value,
and afterwards by a simple periodic sine function. The resulting
incidence counts differ substantially from the typical patterns in
Figure \ref{Figure5}. The importance of such external drive, which
is a crucial aspect of vector transmitted diseases, has been
neglected in most of analyzed models with time and space dependence.

\begin{figure}[!h]
\begin{center}
\includegraphics*[width=7.0cm]{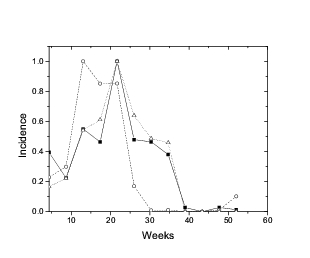}
\end{center}
\caption{\label{Figure6} The normalized rain intensity ($\Gamma_R$),
actual ($I_D$) and simulated ($IH_N$) incidence (by month) in Mossor\'o in
1999. The normalization factors are 71 cases, 149.5 mm, and 72 cases
respectively. Consider the best sample and the following parameter values:
$t_e=4$, $t_l=7$, $t_p=7$ , $t_{lm}=6$, $t_{lh}=5$, $t_v=6$,
$p_{ehm}=p_{emh}=0.75$, $p_{md}=0.143$, $\mu=1.0 \times 10^{-6}$. The used
lines-symbols are, respectively, dashed white-circle, solid black-square, and
dotted grey-triangle.}
\end{figure}

The importance of seasonal aspects for the observed dynamics can be
further exemplified by running the model with the data of other urban
centers. For instance, we consider the 1999 dengue epidemics in Mossor\'o,
in Northeast Brazil \cite{juarez}, for which rainfall peak precedes
$\Gamma_R$ the reported incidence $I_D$ peak. In this case, for which only
monthly data are available for both incidence and rainfall, not only the
rainfall regime is different from that in Salvador, but also notice a
smaller Pearson's correlation coefficient ($c$=$0.69$) between rainfall
and dengue incidence (see Figure \ref{Figure6}).

Mossor\'o's larger  surface of $A_c=2110 \times 10^6$ m \cite{IBGE}
directly influences spatial parameters, leading to a lattice size
$L=554$. As this incidence data is not georeferenced, $\mu$ could
not be directly estimated. However, taking into account that the
city is a less developed urban center with a smaller population than
Salvador ($p_c=$234.390 habitants \cite{IBGE}), we consider a
smaller value of $\mu=1.0 \times 10^{-6}$. The values of other
parameters were chosen according to the already discussed
procedures. We observe that the normalization factors for the actual
(71) and the simulated (72) incidence time series are very similar.
The results in Figures \ref{Figure4} and \ref{Figure6} show that the
model is robust enough to simulate dengue incidence for cities with
high and low rain intensities, and different Pearson correlation
coefficients. Thus, such results indicates that, besides the
importance of periodic forcing, the epidemic behavior of vector
transmitted diseases are heavily dependent on entomological and
epidemiological aspects that are also caught by the model.

\begin{figure}[!h]
\begin{center}
\includegraphics*[width=7.0cm]{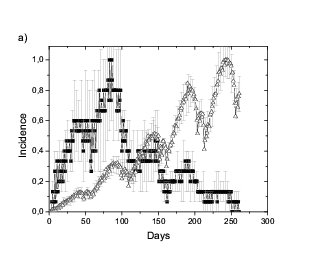}
\includegraphics*[width=7.0cm]{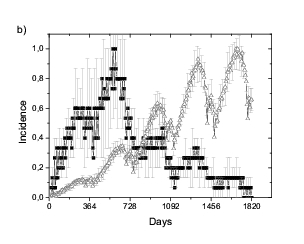}
\end{center}
\caption{\label{Figure7} Predicted average $IH_N$ for a large time
interval of 5 years as function of entomological features. a) Different
values of probability of mosquito death: black squares and dark-grey
triangles indicate, respectively, $p_{md}=1/7=0.143$ and
$p_{md}=1/8=0.125$, while the corresponding normalization factors are 15
and 202. b) Different values of human viremia period: black squares and
dark-grey triangles indicate, respectively, $t_v=5$ and $t_v=7$, with
normalization factors 15 and 219. Other parameter values are the same as
in Figure \ref{Figure4}. Averages and respective error bars (grey)
have been taken over $M_{samples}=20$. }
\end{figure}

To better understand the forcing effect, the behavior of CA model has been
followed for large time intervals. We consider that the exactly the same
rainfall incidence obtained from one-year pluviometric data is repeated
periodically \cite{Vecchio}. Our results indicate the that periodic
forcing leads to modulated responses. However, if we disallow the
possibility of new exogenous infected sources (due, e.g., to an infected
visitor), the amplitude of the epidemic outbursts does not remain the
same. If the same parameter values as in Figure \ref{Figure4} are used,
the results in Figure \ref{Figure7} indicate that $IH_N$ oscillation
amplitude reaches its maximum value in the second year, when it starts
decreasing in a steady way. It is interesting to note that, at the same
time, the $M$ and $A$ populations do not decrease in a similar way. This
indicates that, in a closed environment, the number of individuals
carrying active virus and a relatively weak screening effect due to a
small $RH$ population, turns it difficult to trigger new epidemic events.
Note that, after five years, the number of susceptible individuals $SH$ in
the population is still very high: 97\% for the parameter set that causes
the incidence go to zero.

 On the other hand, Figure \ref{Figure7} also shows that
changes in the parameter values, favoring virus permanence in $M$ and $H$
levels for a longer time, may lead to the opposite landscape, with a long
period during which the yearly amplitude of $IH_N$ population increases
monotonically. In such cases, the amplitude decreases only when a large
fraction of the $H$ population has become infected and switched to the
$RH$ state. Note that this is not yet the situation, after 5 years
evolution period, for such alternative time evolution scenarios. There we
still find a large fraction of $SH$ susceptible individuals: 52\% (see
Figure \ref{Figure7}a), where we introduce a variation of probability of
mosquito death that is the inverse of expected life time of mosquito
($p_{md}=1/8=0.125$), and 46\% (see Figure \ref{Figure7}b) where the
variation occurs in the human viremia ($t_v=7$). The different values of
normalized factors in both cases indicate how these parameters increases
the number of $IH_N$.

This dramatic dependence of the size of successive epidemic events in
isolated environments turns to to be a unexpected result of our model. As
far as we know, this effect, resulting from a local interaction between
the three CA levels, has not been previously discussed in the literature.

\subsection{The mobility effects: spatiotemporal patterns and vector control}
\label{subsec42}

Spatiotemporal patterns resulting from geo-referenced data of the
actuald epidemics of Salvador in 1995 have been reported elsewhere
\cite{Barreto}. They can be compared to the CA simulated
spatiotemporal patterns, which have been generated with the help of
the G2 graphic package \cite{g2_manual}. To this purpose, it is
necessary to assume that, in each CA level, more than one individual
can live in each lattice site. We consider that the total population
of the city is represented by the CA cells, assuming the inhabitants
are a gaussian distributed among the cells with a mean value of 50
humans per cell. With this assumption, the model is able to
reproduce qualitatively the main features observed in actual
spatiotemporal epidemics patterns \cite{Barreto}.

\begin{figure}[!h]
\begin{center}
\includegraphics[width=10cm,angle=0]{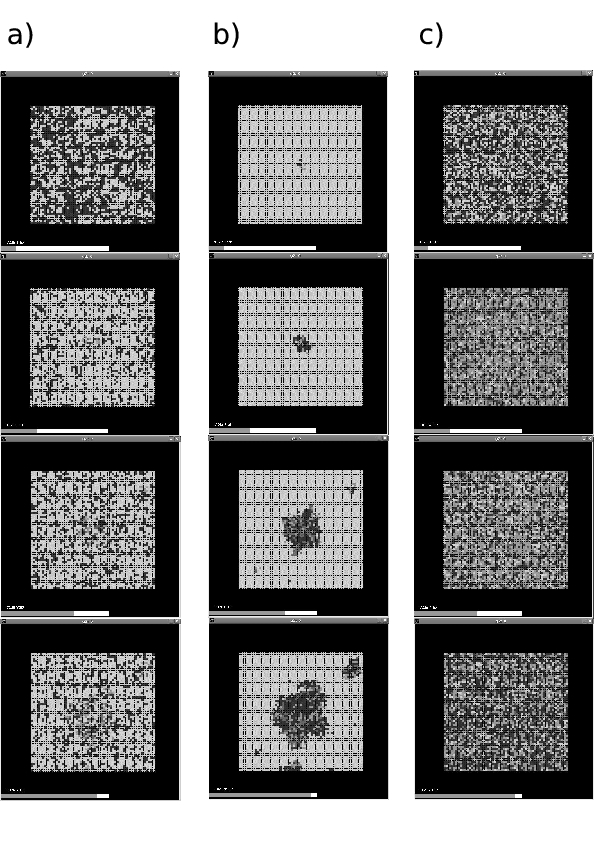}
\end{center}
\caption{\label{Figure8} Simulated spatiotemporal configurations of
cumulated cases, consider one sample and the following parameter values:
$L=79$ , $p_{md}=0.143$ $p_{emh}=p_{ehm}=0.75$ $\mu=0.001$, $t_e=5$, $t_l=
7$ , $t_p= 3$ , $t_{lm}=7$ , $t_{lh}=5$ , $t_v=5$. Four snap shots for
each lattice: a)
Mosquitoes (\emph{M}); b) Humans (\emph{H}); and c) Aquatic phase
(\emph{A}). For on line version: ((\emph{M}): empty site - blue, \emph{SM}
- green, $EM$ - grey, \emph{IM} - red), ((\emph{H}): \emph{SH} - green,
\emph{EH} - grey, \emph{IH} - red, \emph{RH} - blue), ((\emph{A}): empty
site - blue, \emph{E} - green, \emph{L/P} - grey, \emph{B} - red). For
printed version: ((\emph{M}): empty site - white, \emph{SM} - light-grey,
$EM$ - dark-grey, \emph{IM} - black), ((\emph{H}): \emph{SH} - white,
\emph{EH} - light-grey, \emph{IH} - dark-grey, \emph{RH} - black),
((\emph{A}): empty site - white, \emph{E} - light-grey, \emph{L/}P -
dark-grey, \emph{B} - black)}
\end{figure}

In Figure \ref{Figure8}, we illustrate spatiotemporal patterns for
$A$, $M$ and $H$ populations in characteristic time steps. For the
sake of a better visualization, we choose a small value of lattice
size ($L=79$). As initial condition, we assume an infection seed,
represented by one $IH$ site in the $H$ level. Further, due to a
previous large rainfall event, the $E$ and $SM$ states of the $A$
and $M$ levels are largely populated. From this time on, epidemics
starts around the site where the seed was located.  $SM$ changes
into $EM$ state, disseminating the disease into other $H$ sites,
while increasing the radius of the primary epicenter. Due to $H$ and
$M$ mobility, some secondary epicenters are formed. In this case,
without any control strategy, the epidemics evolves naturally until
its end. Figure \ref{Figure8} reveals qualitative similarities to
the main features presented in \cite{Barreto}: the persistence of
the epicenter of the epidemics, the emergence of secondary
epicenters, and an irregular shape of each epicenter.

\begin{figure}[!h]
\begin{center}
\includegraphics[width=7cm,angle=0]{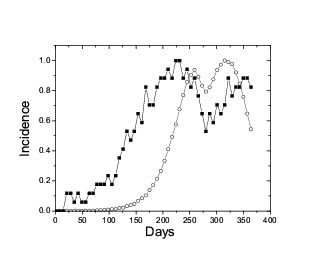}
\end{center}
\caption{\label{Figure9} The mobility parameter effect: normalized $IH_N$
assuming $M_{samples}=200$ and the same parameters values of figure
\ref{Figure4} except $\mu$ that is assumed the following values: 0.0
(black square) and 0.02 (white circle). The normalization factors are,
respectively, 17 and 1867.}
\end{figure}

Secondary epicenters at large distances from the original seed are a
direct consequence of the mobility effects, which are well accepted
to be an important feature for dengue transmission urban centers.
Indeed, if $\mu=0$, the shown spatiotemporal pattern is replaced by
a diffusion-like pattern with a single epicenter. However, $\mu$
also plays an important in reducing time series fluctuations, an
expected `mean-field' effect related to the global infection
probability. This effect is made clear in Figure \ref{Figure9}. The
curves also show that non-zero values of $\mu$ introduce a time
delay effect extending the duration and the intensity of the
epidemics process. Indeed, the large difference in the normalization
factors for both curves indicates that $\mu$ is directly related to
a much faster epidemic dissemination.

Until today, no efficient vaccine against dengue could be devised.
Therefore, actions towards vector control constitute the only public
health policy to reduce the deleterious effect of the disease. Even so,
there are still controversies regarding whether vector control actions are
more reliable in the $A$ or $M$ phases. As the CA model is able to
successfully reproduce epidemics data and follow the dynamics of the
disease for longer periods of time, it can also provide useful insights
regarding the effect produced by different vector control mechanisms.

To this purpose, let us consider the dependence between the epidemic
threshold and the vector control parameter $p_{amd}$. We have performed a
large number of independent simulations for different values of $p_{amd}$.
We evaluated $I_{th}$ with the help of equation (\ref{epidemics}), where
the time average was replaced by sample averages. Thus, $I_{th}$ is
directly related to the probability that an individual living the the
simulated urban center gets infected within a one-year time span.

\begin{figure}[!h]
\begin{center}
\includegraphics[width=7cm,angle=0]{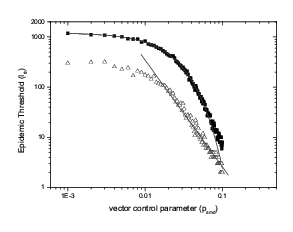}
\end{center}
\caption{\label{Figure10} The vector control analysis: the simulated
$I_{th} \times$ vector control parameter ($p_{amd}$). We consider
$M_{samples}=20$ and the same parameters values of Figure \ref{Figure4},
except for the parameter $\mu=0$ and $\mu=10^{-5}$. The corresponding
values are indicated by, respectively, white triangles and black squares.
The parameter of the power law fitting (grey), for $\mu=0$, is the
exponent $\alpha =-2.13 \pm 0.04$ and $a=-1.72 \pm 0.05$. The parameters
of the stretched exponential fitting (grey), for $\mu = 10^{-5}$, are
$b_1=1218.04 \pm 7.00$, $b_2=86.44 \pm 4.00$, and the exponent $\beta=1.14
\pm 0.01$.}
\end{figure}

The results in Figure \ref{Figure10} show that, when $\mu=0$, the
dependence between $I_{th}$ and $p_{amd}$  follow a power law
behavior, $I_{th} = a p_{amd}^\alpha$, with large values. It clearly
shows that effective policies aiming at a reduction of the vector
reproduction in its own environment produce substantial reduction of
affected population. This effect is still more expressive and
relevant when we consider more realistic situations, in which human
and vector population move in the urban space. Indeed, when $\mu >
0$, $I_{th}$ decays with respect to $p_{amd}$ in a faster way the
points of fit quite well to a stretched exponential $I_{th}=b_1
\exp[-b_2(p_{amd}^{\beta})]$. Moreover, as expected, the epidemic
threshold is larger, for any value of $p_{amd}$, when $\mu
>0$ than when $\mu =0$.

\section{Concluding Remarks and Perspectives}
\label{sec5}

The three level CA model investigated in this work presents several
features that allow for a quantitative reproduction of actual time series
of dengue epidemics. Besides the usual local interaction steps based on
SEIR compartment models, the most important novelties are: i) the use of
the climatical data as input data; ii) the $A-M$ and $M-H$ inter-level
interactions; iii) the inclusion of short-range vector mobility and
long-range human mobility.

The model is robust with respect to the range of parameters
considered in the literature, and to its ability in reproducing time
series of dengue epidemics in different urban centers. The climatic
input data as well as the procedure used for estimating the
parameter values are able to catch the diversity of the time series
dengue incidence for different cities. Although we have mainly
focused our analysis on the human population, the CA model also
provides useful insights on the behavior of the vector population,
which will be presented in a future work.

The effect of periodic forcing allows us to suggest effective measures to
reduce the probability of recurrent outbursts. Indeed, the effect of an
increased infected vector life time is found to be very important to alter
of the magnitude of epidemic events.

The analysis of vector control shows that, as expected, it indeed
produces a decrease in the probability of human infection. However,
we have shown that this effect is more relevant when vector and
human mobility are taken into account. In this case, the infection
probability decreases according to an stretched exponential, while a
power law behavior is observed when the no mobility assumption is
taken into account.

Perspectives for further work on this model are of two kind. The first one
amounts to investigate the the impact of different strategies of vector
control on dengue transmission as well as to discuss the detailed behavior
of $M$ and $A$ populations subject to those strategies. A more ambitions
goal is to achieve the quantitative reproduction of spatial patterns. This
requires a more precise local characterization of spatial units, as well
as a more precise GPS georeferencing data. This way, the CA model can help
to plan improved vector control policies from the spatial point of view,
attacking mainly the most important focus for the propagation of the
epidemics. \vspace{2cm}

{\bf Acknowledgements:} The authors thank C. P. Ferreira, D. Alves,
E. Massad, H. M. Yang, J. G. V. Miranda, J. P. Dias, L. Esteva, M.
N. Burattini, V. C. G. S. Morato  for useful discussions about
dengue modelingd. The authors acknowledge the Brazilian agencies
CNPq and FAPESB for financial support.

\end{document}